\documentclass[12pt]{article}

\usepackage{epsfig} 
\usepackage{epstopdf}
\usepackage{amssymb}
\usepackage{amsmath}
\usepackage{amsfonts}
\usepackage{graphicx}
\graphicspath{ {plots/} }
\usepackage{mathrsfs}
\usepackage{color}
\usepackage{multirow}
\usepackage{psfrag}

\newcommand{\C}{\mathbb{C}}

\newcommand{\fa}{\mathfrak{a}}
\newcommand{\fb}{\mathfrak{b}}
\newcommand{\fc}{\mathfrak{c}}

\newcommand{\fu}{\mathfrak{u}}

\newcommand{\fn}{{\mathfrak{n}}}

\newcommand{\fz}{\mathfrak{z}}

\newcommand{\fK}{\mathfrak{K}}

\newcommand{\bz}{\mathbf{z}}

\newcommand{\cO}{\mathcal{O}}
\newcommand{\cP}{\mathcal{P}}

\newcommand{\cT}{\mathcal{T}}

\newcommand{\cX}{\mathcal{X}}
\newcommand{\cY}{\mathcal{Y}}
\newcommand{\be}{\begin{equation}}
\newcommand{\ee}{\end{equation}}
\newcommand{\bea}{\begin{eqnarray}}
\newcommand{\eea}{\end{eqnarray}}
\newcommand{\nn}{\nonumber}

\newcommand{\ed}{\end{document}}

\newcommand{\rC}{{\rm C}}
\newcommand{\rS}{{\rm S}}

\newcommand{\bi}{\begin{itemize}}
\newcommand{\ei}{\end{itemize}}

\newcommand{\bce}{\begin{center}}
\newcommand{\ece}{\end{center}}

\newcommand{\sE}{\mathscr{E}}

\newcommand{\sH}{\mathscr{H}}

\newcommand{\RE}{{\rm Re}}
\newcommand{\IM}{{\rm Im}}

\begin{document}

\title{Nonlinear Spectral Singularities and Laser Output Intensity}

\author{Hamed Ghaemi-Dizicheh$^1$, Ali~Mostafazadeh$^{1,2,}$\thanks{Corresponding author, Email Address: amostafazadeh@ku.edu.tr} , and Mustafa Sar{\i}saman$^1$\\[6pt]
Departments of Physics$^1$ and Mathematics$^2$, Ko\c{c} University,\\ 34450 Sar{\i}yer,
Istanbul, Turkey}

\date{ }
\maketitle

\begin{abstract}

The mathematical notion of spectral singularity admits a description in terms of purely outgoing solutions of a corresponding linear wave equation. This leads to a nonlinear generalization of this notion for nonlinearities that are confined in space. We examine the nonlinear spectral singularities in arbitrary TE and TM modes of a mirrorless slab laser that involves a weak Kerr nonlinearity. This provides a computational scheme for the determination of the laser output intensity $I$ for these modes. In particular, we offer an essentially mathematical derivation of the linear-dependence of $I$ on the gain coefficient $g$ and obtain an explicit analytic expression for its slope. This shows that if the real part $\eta$ of the refractive index of the slab does not exceed 3, there is a lower bound on $\theta$ below which lasing in both its TE and TM modes requires $\eta$ to be shifted by a small amount as $g$ surpasses the threshold gain. Our results suggest that lasing in the oblique TM modes of the slab is forbidden if the incidence (emission) angle of the TM mode exceeds the Brewster's angle.
\vspace{2mm}



\end{abstract}

\section{Introduction}

A planar slab does not reflect the transverse magnetic waves with an incident angle $\theta$ equal to Brewster's angle $\theta_b$, \cite{BW}. This well-known observation suggests that it should be extremely difficult for a mirrorless slab laser to emit a TM wave at the Brewster's angle, because the internal reflections, whose number determines the optical path of the wave inside the slab, are essentially absent for such a wave. Ref.~\cite{pra-2015a} establishes the validity of this prediction by deriving explicit formulas for the laser threshold condition governing arbitrary transverse electric (TE) and transverse magnetic (TM) modes of a homogeneous slab of gain material.
    \begin{figure}
    \begin{center}
    \includegraphics[scale=.45]{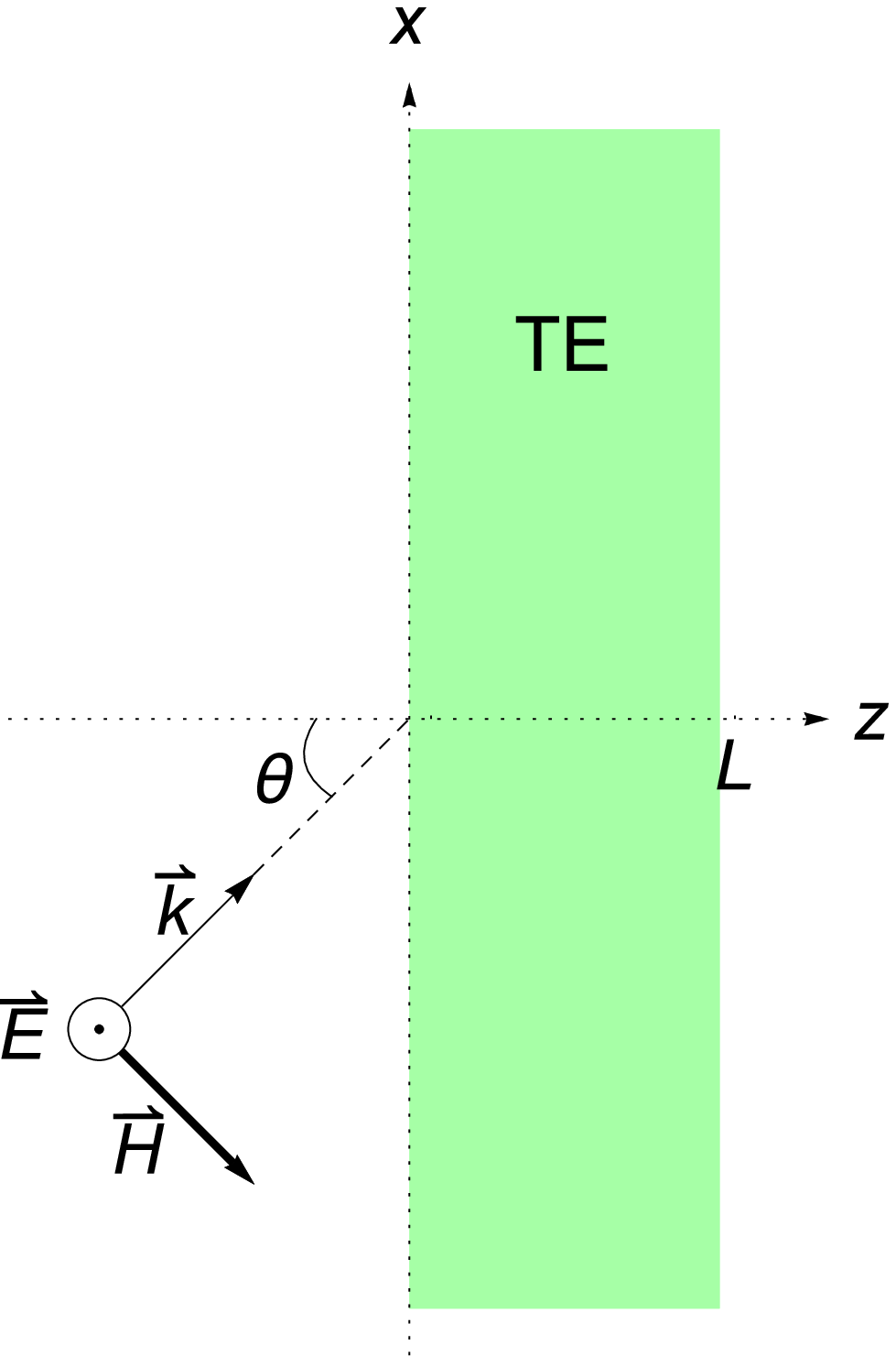}~~\includegraphics[scale=0.45]{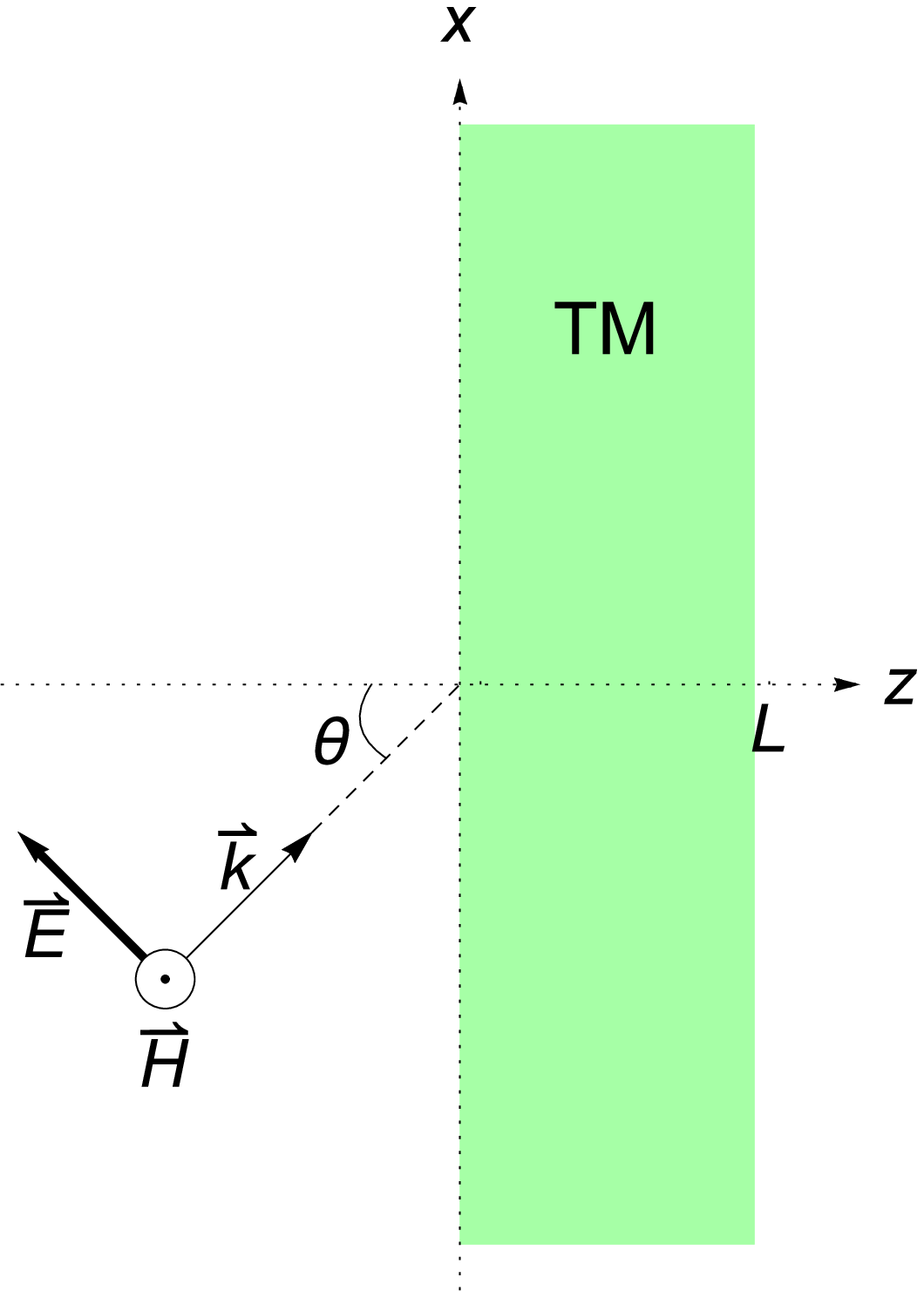}
    \caption{(Color online) Schematic representation of the TE (on the left) and TM (on the right) modes of a planar slab of thickness $L$ that is made out of a homogeneous gain material.}
    \label{fig1}
    \end{center}
    \end{figure}
A closer look at the behavior of the singular waves, i.e., those generated at the threshold gain, reveals unexpected differences between the TM waves with $\theta<\theta_b$ and $\theta>\theta_b$. For example, the energy density of these waves turns out to take larger values outside the slab for $\theta>\theta_b$, while the converse is the case for TM waves with $\theta<\theta_b$ and TE waves with arbitrary $\theta$. This observation will have physical significance, only if it applies also for the cases that the gain coefficient of the slab exceeds its threshold value so that the slab actually emits laser light. The purpose of the present article is to address this problem by offering a comprehensive analysis of the laser output intensity for general TE and TM modes of the slab laser.

The approach of \cite{pra-2015a} is based on the observation that the laser threshold condition can be derived as the condition for the presence of a spectral singularity \cite{prl-2009}. This is originally demonstrated for the normally incident TE modes of a homogeneous slab in \cite{pra-2011a}, and subsequently used to derive the laser threshold condition for active media of various geometries \cite{ss1-ss2}. The first attempt at using this method for the description of obliquely incident TE modes of a homogeneous slab is due to Aalipour \cite{aalipour} whose results are improved and generalized to the TM waves in \cite{pra-2015a}. For other studies of physical aspects of spectral singularities, see \cite{Longhi,CP-2012,GC-2013-14,HR,Li-2014,wang-2016,Hang-2016,Kalozoumis-2016,Pendharker-2016,jiang-2016}.

A related recent development is the introduction of a nonlinear generalization of the concept of spectral singularity \cite{prl-2013}. This turns out to offer a simple effective description of the linear dependence of the laser output intensity on the gain coefficient  \cite{Silfvast}. Here the idea is to postulate the existence of a weak Kerr nonlinearity in the permittivity of the medium and use the condition for the emergence of a nonlinear spectral singularity to derive an expression for the intensity $I$ of the emitted waves. For both homogeneous \cite{pra-2013c} and $\cP\cT$ symmetric bilayer slabs \cite{sam-2014}, this scheme yields
    \be
    I=\left(\frac{g-g_0}{\sigma g}\right)\widehat I,
    \label{eq1}
    \ee
where $g$ is the gain coefficient of the active component of the system, $g_0$ is its threshold value, $\sigma$ is the Kerr coefficient, and $\widehat I$ is a function of the geometry and other parameters of the system. For other developments related to the optical applications of nonlinear spectral singularities, see \cite{reddy,liu}.

In the present article, we use nonlinear spectral singularities to offer a derivation of (\ref{eq1}) for general TE and TM modes of a homogeneous slab. In particular, we give an explicit formula for $\widehat I$ and explore its physical consequences.

\section{TE modes}
\label{TE}

Consider time-harmonic electromagnetic waves, $e^{-i\omega t}{\vec E}({\vec r})$ and $e^{-i\omega t}{\vec H}({\vec r})$, interacting with an infinite homogeneous and isotropic planar slab of thickness $L$ that is aligned in the $x$-$y$ plane. The inclusion of a Kerr nonlinearity amounts to having the permittivity of the slab given by
    \be
    \epsilon({\vec r})=\epsilon_0[\fn^2+\sigma|{\vec E}({\vec r})|^2],
    \label{kerr}
    \ee
where $\epsilon_0$ is the permittivity of the vacuum, $\fn$ is the complex refractive index of the slab in the absence of the nonlinearity, and $\sigma$ is the Kerr coefficient \cite{boyd}.

For the TE waves, ${\vec E}({\vec r})$ is parallel to the faces of the slab. We choose a coordinate system in which it is aligned along the $y$-axis. We can then write ${\vec E}({\vec r})$ and the wavevector $\vec k$ in the form
    \begin{align}
    &{\vec E}({\vec r})=e^{ik_{x}x} \sE(z){\vec e}_y, &&\vec k=k_x{\vec e}_x+k_z{\vec e}_z,
    \label{TE-defn}
    \end{align}
where ${\vec e}_x$, ${\vec e}_y$, and ${\vec e}_z$ are respectively the unit vectors pointing along the positive $x$-, $y$-, and $z$-axes,
    \begin{align}
    &k_x:=k\sin\theta, &&k_z:=k\cos\theta,
    \label{kx-kz=}
    \end{align}
$k:=\omega/c$ is the wavenumber, $\theta$ is the incidence angle depicted in Fig.~\ref{fig1}, and $\sE(z)$ is a function whose form is determined by Maxwell's equations \cite{jackson}. In view of (\ref{kerr}) -- (\ref{kx-kz=}), these reduce to
    \begin{align}
    &{\vec H}({{\vec r}}) = i (k Z_{0})^{-1}e^{ik_x x}  \left[\sE'(z){\vec e}_x-ik_x\sE(z){\vec e}_z\right],
    \label{TE-H=}\\[3pt]
    &\sE''(z) +k^2[\hat\epsilon(z)-\sin^2\theta]\sE(z) = 0,
    \label{TE-E=}
    \end{align}
where
    \be
    \hat\epsilon(z):=\left\{\begin{array}{cc}
    \fn^2+\sigma|\sE(z)|^2 & {\rm for}~0\leq z\leq L,\\[3pt]
    1 & {\rm otherwise},
    \end{array}\right.
    \label{e1}
    \ee
is the relative permittivity.

Let us introduce:
    \be
    \begin{aligned}
    &\mathbf{z}:=\frac{z}{L}, &&\fK:=Lk_z=kL\cos\theta\\
    &\gamma:=-\sigma k^2L^2,
    &&\tilde\fn:=\sec\theta\sqrt{\fn^2-\sin^2\theta}.
    \end{aligned}
    \label{scaled}
    \ee
Then according to (\ref{TE-E=}) and (\ref{e1}),
    \be
    \sE(L\bz)=\left\{\begin{array}{ccc}
    A_{-}e^{i\fK\,  {\bz}}+B_{-}e^{-i\fK\, {\bz}}&{\rm for}& {\bz}< 0,\\[3pt]
    \zeta(\bz)&{\rm for}& {\bz}\in [0,1],\\[3pt]
    A_{+}e^{i\fK\, {\bz}}+B_{+}e^{-i\fK\, {\bz}}&{\rm for}& {\bz}> 1.
    \end{array}\right.
    \label{e2}
    \ee
where $\zeta:[0,1]\to\C$ satisfies the nonlinear Schr\"odinger equation:
    \be
    -\zeta''(\mathbf{z})+\fK^{2}(1-\tilde\fn^2)\zeta(\mathbf{z})+\gamma \vert\zeta(\mathbf{z})\vert^2\zeta(\mathbf{z})=\fK^{2}\zeta(\mathbf{z}),
    \label{zeta-eqn}
    \ee
for $\mathbf{z}\in(0,1)$.

Each solution of (\ref{zeta-eqn}) determines a TE wave interacting with the slab provided that we impose appropriate boundary condition that fix $\zeta(0)$ and $\zeta(1)$. These follow from the standard boundary conditions fulfilled by the electric and magnetic fields \cite{jackson}, i.e., their tangential component must be continuous at the interfaces. In light of (\ref{TE-defn}) and (\ref{TE-H=}), this implies that $\sE(z)$ and $\sE'(z)$ are continuous functions at $z=0$ and $L$. Therefore, we can apply the general formalism developed in Ref.~\cite{prl-2013} to describe the scattering of TE waves. In particular, the reflection and transmission amplitudes, $R^{\rm l}$ and $T^{\rm l}$, for the left-incident TE waves are given by
    \begin{align}
    &R^l=-\frac{G_-(\fK)}{G_+(\fK)}, &&T^l=\frac{2i\fK N_+}{G_+(\fK)},
    \label{R-T}
    \end{align}
where
    \be
    G_\pm(\fK):=\zeta'(0)\pm i\fK\,\zeta(0),
    \label{Gp=}
    \ee
and $N_+$ is the complex amplitude of the transmitted wave, so that $\sE(z)=N_+ e^{ik_{z}{z}}$ for $z\geq L$. This together with (\ref{e2}) imply
    \be
    \zeta(1)=\frac{\zeta'(1)}{i\fK}=N_+ e^{i\fK}.
    \label{BC-TE=}
    \ee
Nonlinear spectral singularities correspond to real and positive values of $\fK$ for which
    \be
    G_+(\fK)=0.
    \label{Gp=0}
    \ee

Assuming that the strength of the nonlinearity is so small that we can ignore quadratic and higher order terms in $\gamma$, we can attempt to solve $(\ref{Gp=0})$ using first-order perturbation theory. For $\theta=0$, i.e., normally incident TE waves, this has been done in Ref.~\cite{pra-2013c}. It is easy to see that setting $\theta=0$ in (\ref{zeta-eqn}) amounts to changing $\tilde\fn$ to $\fn$. This in turn suggests that we can determine both the linear and nonlinear spectral singularities in arbitrary TE modes of our slab by replacing $\fn$ by $\tilde\fn$ in the analysis of the normally incident TE waves that is offered in Refs.~\cite{pra-2011a,pra-2013c}.

Using $\fn_0$ and $k_0$ to identify the values of $\fn$ and $k$ that yield a linear spectral singularity, introducing
    \begin{align}
    &\tilde\fn_0:=\sec\theta\sqrt{\fn_0^2-\sin^2\theta}, && \fK_0=k_0L\cos\theta,
    \label{zero-symb-1}
    \end{align}
and substituting $\tilde\fn$ for $\fn$ in Eq.~(8) of Ref.~\cite{pra-2011a}, we find
    \be
    e^{-i\tilde{\fn}_{0}\fK_{0}}= \frac{\tilde{\fn}_{0}-1}{\tilde{\fn}_{0}+1}.
    \label{linear-SS}
    \ee
Next, we denote the real and imaginary parts of $\fn$ (respectively $\fn_0$) by $\eta$ and $\kappa$ (respectively $\eta_0$ and $\kappa_0$), so that
    \begin{align}
    &\fn=\eta+i\kappa,
    &&\fn_0=\eta_0+i\kappa_0,
    \label{eta-kappa-zero}
    \end{align}
and recall that the gain coefficient $g$ of the slab is given by \cite{Silfvast}:
    \be
    g=-2k\kappa.
    \label{g-zero}
    \ee
The threshold gain $g_0$ is the value of $g$ for $k=k_0$ and $\kappa=\kappa_0$. We can determine $g_0$ and $k_0$ by equating the absolute-value and the phase of the left- and right-hand sides of (\ref{linear-SS}), respectively, \cite{pra-2015a}. This gives the following expressions for the  threshold gain and the wavenumber of the emitted wave:
    \bea
    g_0&=&\frac{2\,\IM(\fn)}{L\cos\theta\,\IM(\tilde\fn)}
    \ln\left|\frac{\tilde\fn+1}{\tilde\fn-1}\right|,
    \label{th-g=1}\\
    k_0&=&\frac{\pi m-\varphi_0}{L\cos\theta\,\RE(\tilde\fn)},
    \label{th-lambda=1}
    \eea
where $m$ is a positive integer (mode number) and $\varphi_0$ is the phase angle (principal argument) of $(\tilde\fn-1)/(\tilde\fn+1)$, \cite{footnote1}.

\begin{figure}
	\begin{center}
	\includegraphics[scale=.60]{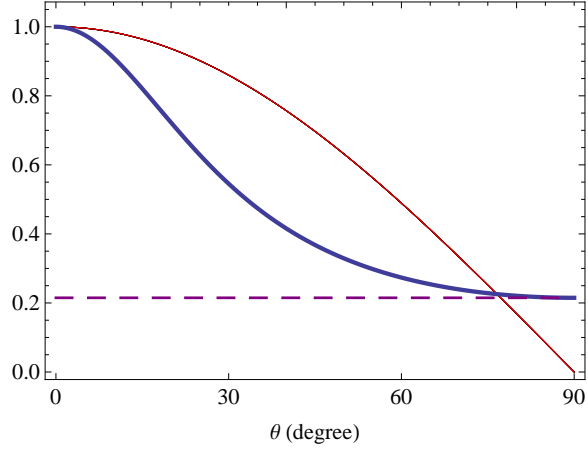}
	\caption{Plot of $g_0(\theta)/g_0(0^\circ)$ (thin solid red curve) and $\widehat I(\theta)/\widehat I(0^\circ)$ (thick solid navy curve) for $\eta_0=3.4$, $L=300~\mu{\rm m}$, and $\lambda=1500~{\rm nm}$. Here $g_0(0^\circ)=40.409~{\rm cm}^{-1}$ and $\widehat I(0^\circ)=393.089$. For $\theta\to 90^\circ$, $\widehat I(\theta)$ tends to $84.480$. The dashed (purple) line represents the corresponding asymptote.}
	\label{fig2}
	\end{center}
	\end{figure}

Suppose that our slab is made of a typical high-gain (nongaseous) active material and that its
thickness is much larger than the wavelength of the emitted wave, so that
    \be
    |\kappa_0|\ll \eta_0-1\ll k_0L.
    \label{typical}
    \ee
Then (\ref{th-g=1}) and (\ref{th-lambda=1}) give
    \bea
    g_0&\approx&\frac{4\sqrt{\eta_0^2-\sin^2\theta}}{L\eta_0}\ln
    \frac{|\sqrt{\eta_0^2-\sin^2\theta}+\cos\theta|}{\sqrt{\eta_0^2-1}},
    \label{threshold-g}\\
    k_0&\approx&\frac{\pi m}{L\sqrt{\eta_0^2-\sin^2\theta}},
    \label{k-zero=}
    \eea
where `$\approx$' labels approximate equalities in which we neglect the first and higher order terms in $\kappa_0/m$ and the quadratic and higher order terms in powers of $\kappa_0$. Fig.~\ref{fig2} shows a plot of $g_0$ as a function of $\theta$ for $L=300~\mu{\rm m}$ and $\eta_0=3.4$. For the wavelength $\lambda_0:=2\pi/k_0=1500~{\rm nm}$, which is possible for  $1300\leq m\leq 1360$, the approximate expression (\ref{threshold-g}) for $g_0$ yields values that agree with the exact (numerical) values obtained directly using (\ref{linear-SS}) to 6 significant figures.

Following a similar approach we can characterize the nonlinear spectral singularities in the oblique TE modes of the slab from those in its normally incident TE mode. To do this we express the values of $\fn$ and $k$ that give rise to a nonlinear spectral singularities in the form
    \begin{align}
    &\fn=\fn_0+\gamma\fn_1, &&k=k_0+\gamma k_1,
    \label{1st-order-n-k}
    \end{align}
where $\fn_1$ and $k_1$ are respectively complex and real parameters. For a sufficiently weak nonlinearity, the $\gamma$-dependence of $\fn_1$ and $k_1$ is negligible. This suggests that we treat $\gamma$ as a perturbation parameter and use first-order perturbation theory to compute $\zeta(\bz)$ and $G_+(k)$, and solve (\ref{Gp=0}). This is done in Ref.~\cite{pra-2013c} for $\theta=0$. We can treat (\ref{Gp=0}) for the case $\theta\neq 0$ by setting $\fn_0\to\tilde\fn_0$ and
    \begin{align}
    &\fK_0:=Lk_0\cos\theta, && \fK_1:=Lk_1\cos\theta,
    \label{set}
    \end{align}
in Eq.~(31) of \cite{pra-2013c}. This gives the following relation for the existence of a nonlinear spectral singularity.
   \be
    \fK_1=\fa \fn_1+\fb I,
    \label{NL-SS-TE-1}
    \ee
where $I:=|N_+|^2/2$ is the time-averaged intensity of the emitted wave from the right, and
    \bea
    \fa&:=&\frac{-\fn_0[(\tilde\fn_0^2-1)\fK_0-2i]\sec^2\theta}{\tilde\fn_0^2(\tilde\fn_0^2-1)},
    \label{a=}\\
    \fb&:=&\frac{-16i(4\tilde\fn_0^2-\tilde\fn_0^{*2}-3)}{
    \fK_0^2(\tilde\fn_0^2-1)(9\tilde\fn_0^4+\tilde\fn_0^{*4}-10|\tilde\fn_0|^4)}.
    \label{b=}
    \eea

Next, we introduce
    \begin{align}
    & \eta_1:=\RE(\fn_1),  && \kappa_1:=\IM(\fn_1), && \fa_r:=\RE(\fa),
    \label{eta1=}\\
    & \fa_i:=\IM(\fa), && \fb_r:=\RE(\fb), && \fb_i:=\IM(\fb),
    \end{align}
where `$\RE$' and `$\IM$' stand for the real and imaginary part of their argument, respectively. Then,
	\begin{align}
    	&\fn_1=\eta_1+i\kappa_1, &&\fa=\fa_r+i\fa_i, && \fb=\fb_r+i\fb_i,
    \label{new-symb1}
	\end{align}
and (\ref{NL-SS-TE-1}) is equivalent to
	\bea
	\fK_1&=&\fa_r^{-1}|\fa|^2\eta_1+(\fb_r+\fa_r^{-1}\fa_i\fb_i)I,
	\label{fK-1=}\\
	\kappa_1&=&-\fa_r^{-1}(\fa_i\eta_1+\fb_i I).
	\label{kappa-1=}
	\eea
These relations describe the nonlinear spectral singularities associated with a weak Kerr nonlinearity in the TE modes of our slab. In order to see how we can use them to relate the laser output intensity to the gain coefficient, we first recall that to the first order in $\gamma$ the gain coefficient is given by \cite{pra-2013c}:
	\be
	g=g_0\left[1+\gamma\left(\frac{\fK_1}{\fK_0}+\frac{\kappa_1}{\kappa_0}\right)\right].
	\label{g-g0}
	\ee
This is a simple consequence of (\ref{g-zero}).
	
If we assume that the presence of the nonlinearity does not change the value of the refractive index $\fn$,  we have $\eta_1=0$, and (\ref{fK-1=}) and (\ref{kappa-1=}) reduce to
	\begin{align}
	&\fK_1=(\fb_r+\fa_r^{-1}\fa_i\fb_i)I,
	&& \kappa_1=-\fa_r^{-1}\fb_i I.
	\label{kappa-1-fK_1=}
	\end{align}
Substituting these relations in (\ref{g-g0}) and making use of (\ref{scaled}), we obtain (\ref{eq1}) with $\widehat I$ given by
	\be
	\widehat I=\frac{\kappa_0 \fa_r  \cos^2\theta}{\fK_0
	\left[\fK_0\fb_i-(\fa_r\fb_r+\fa_i\fb_i)\kappa_0\right]}.
	\label{I-TE=}
	\ee
Next, we expand the right-hand side of this relation in a power series in $\kappa_0$. Then using $\kappa_0\fK_0=-g_0L\cos\theta/2$, (\ref{threshold-g}), (\ref{typical}), and (\ref{I-TE=}), we find
	\be
	\widehat I\approx\frac{8(\eta_0^2-1)(\eta_0^2-\sin^2\theta)^{3/2}}{
    	\cos\theta(\eta_0^2+8\sin^2\theta-9)}
    	\ln \frac{\sqrt{\eta_0^2-\sin^2\theta}+\cos\theta}{\sqrt{\eta_0^2-1}} .
    	\label{I-TE=approx}
	\ee
For $\theta=0$ this relation reduces to
    \be
    \widehat I\approx\frac{4 \eta_0^3 (\eta_0^2-1)}{\eta_0^2-9}\ln\left(\frac{\eta_0+1}{\eta_0-1}\right).
    \label{I-zero}
    \ee
Figure~\ref{fig2} shows the graph of the normalized threshold gain $g_0$ and $\widehat I$ as a function of $\theta$ for $\eta_0=3.4$. We have checked that the graphs obtained using the exact and approximate expression for $g_0$ and $\widehat I$ are indistinguishable. As one increases $\theta$, both $g_0$ and $\widehat I$ decrease monotonically. For $\theta\to 90^\circ$, $g_0$ tends to zero while $\widehat I$ approaches $84.480$.

Equations~(\ref{I-TE=approx}) and (\ref{I-zero}) reveal a surprising behavior; for $\eta_0\leq 3$, the intensity factor $\widehat I$ blows up at $\theta=\theta_\star$, where
    \[\theta_\star\approx \sin^{-1}\sqrt{\frac{9-\eta_0^2}{8}}.\]
Furthermore, $\widehat I>0$ if and only if $\theta>\theta_\star$. These observations are easily validated using exact (numerical) treatments of (\ref{I-TE=}) that do not rely on the condition (\ref{typical}).

The singular behavior of $\widehat I$ at $\theta=\theta_\star$ shows that our first-order perturbative derivation of (\ref{eq1}) with $\widehat I$ given by (\ref{I-TE=}) is inapplicable for $\theta\approx\theta_\star$. The problem with negative values of $\widehat I$ is more serious. We can interpret it in either of the following alternative ways.
    \begin{itemize}
    \item It is a sign of the inapplicability of our derivation of (\ref{eq1}) in terms of nonlinear spectral singularities.
    \item It suggests that a mirrorless slab made of gain material with $\eta_0\leq 3$ does not actually lase in its TE modes with $\theta<\theta_\star$.
    \item It shows that making the gain coefficient exceed its threshold value does change the value of $\eta$, i.e., the assumption that $\eta_1=0$ is false.
    \end{itemize}
The latter interpretation calls for a generalization of (\ref{I-TE=}) that does not rely on the condition: $\eta_1=0$. Inserting (\ref{fK-1=}) and (\ref{kappa-1=}) in (\ref{g-g0}) and making use of (\ref{scaled}), we have
    \be
    I=\left(\frac{g-g_0}{\sigma g_0}\right)\widehat I_0+J\,\eta_1,
    \label{I-TE-3}
    \ee
where $\widehat I_0$ stands for the right-hand side of (\ref{I-TE=}), i.e.,
	\be
	\widehat I_0:=\frac{\kappa_0 \fa_r  \cos^2\theta}{\fK_0
	\left[\fK_0\fb_i-(\fa_r\fb_r+\fa_i\fb_i)\kappa_0\right]},\nn
	\ee
and
    \be
    J:= \frac{|\fa|^2\kappa_0-\fa_i\fK_0}{\fb_i\fK_0-(\fa_r\fb_r+\fa_i\fb_i)\kappa_0}.\nn
    \ee

\begin{figure}
	\begin{center}
	\includegraphics[scale=.60]{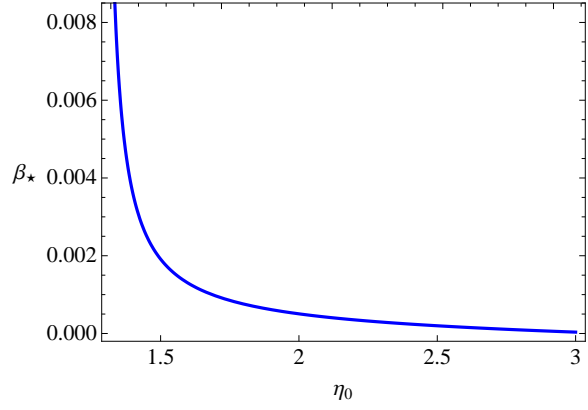}
	\caption{Plot of $\beta_\star$ as a function of $\eta_0$ for $L=300~\mu{\rm m}$ and $\lambda=1500~{\rm nm}$.}
	\label{fig03}
	\end{center}
	\end{figure}

To proceed further we need more information about the behavior of $\eta_1$. Given that the refractive index is in principle a function of the frequency and therefore the wavenumber, we expect that the change in the value of $\fK$, i.e., $\fK_0\to\fK_0+\gamma\fK_1$, is responsible for the change in $\eta$, i.e., $\eta_0\to\eta_0+\gamma\eta_1$. Because $\gamma\ll 1$, this suggests that
	\be
	\eta_1=\beta\,\fK_1,
    	\label{eta-1-beta}
    	\ee
where $\beta$ is a real parameter. Substituting this equation in (\ref{fK-1=}) and solving for $\fK_1$ gives
	\be
	\fK_1=K I,~~~~~~K:=\frac{\fa_r\fb_r+\fa_i\fb_i}{\fa_r-\beta|\fa|^2}.
	\label{K1=301}
	\ee
In view of (\ref{eta-1-beta}) and (\ref{K1=301}), (\ref{I-TE-3}) coincides with (\ref{eq1}) provided that we take
	\bea
    	\widehat I&=&\frac{\widehat I_0}{1-\beta J K}=
    	\frac{\kappa_0 (\fa_r-\beta|\fa|^2)  \cos^2\theta}{\fK_0
	\left\{\fK_0[\fb_i-\beta(\fa_r\fb_i-\fb_r\fa_i)]-(\fa_r\fb_r+\fa_i\fb_i)\kappa_0\right\}}.
	\label{I-TE=gen-beta}
	\eea
A closer examination of this relation shows that for $\eta_0\leq 3$, there is a positive number $\beta_\star$ such that the right-hand side of (\ref{I-TE=gen-beta}) becomes positive for all $\theta\in[0,90^\circ]$ if and only if $\beta>\beta_\star$. For a sample with $\eta_0=2$, $L=300\:\mu{\rm m }$, and $\lambda=1500\:{\rm nm}$, we have $\beta_\star=4.724\times 10^{-4}$. Figure~\ref{fig03} shows a plot of $\beta_\star$ as a function of $\eta_0$. It reveals the
fact that $\beta_\star$ is a monotonically decreasing function of $\eta_0$.

\section{TM modes}
\label{TM}

\subsection{Nonlinear Helmholtz Equation}

For TM waves, we can write $\vec H$ in terms of a scalar function $\sH$ according to
    \bea
    \vec H(\vec r)&=&e^{ik_x x}\sH(z)\vec e_y.
    \label{H-TM}
    \eea
Substituting this relation in Maxwell's equations, we find
    \begin{align}
    &\vec E(\vec r)=\frac{iZ_0 e^{ik_x x}}{k\,\hat\epsilon(z)}
    \left[-\sH'(z)\vec e_x+ik_x\sH(z)\vec e_z\right],
    \label{E-TM}\\
    &\sH''(z)-\hat\epsilon(z)^{-1}\hat\epsilon'(z)\sH'(z)+
    k^2[\hat\epsilon(z)-\sin^2\theta]\sH(z)=0,
    \label{HE-TM}
    \end{align}
where $Z_0:=\sqrt{\mu_0/\epsilon_0}$ and $\mu_0$ are respectively the impedance and permeability of the vacuum. Equation~(\ref{HE-TM}) is to be solved subject to the following matching conditions \cite{jackson}.
    \begin{align}
    &\sH(0^-)=\sH(0^+), &&\sH(L^-)=\sH(L^+),
    \label{H-BC}\\
    &\sH'(0^-)=\frac{\sH'(0^+)}{\hat\epsilon(0)},
    &&\frac{\sH'(L^-)}{\hat\epsilon(L)}=\sH'(L^+),
    \label{dH-BC}
    \end{align}
where $f(z_0^{-/+})$ stands for the left/right limit of $f(z)$ as $z\to z_0$. Notice that because the relative permittivity depends on $\vec E$, there is a nonlinear coupling between Eqs.~(\ref{E-TM}) and (\ref{H-TM}) and the matching conditions (\ref{H-BC}) and (\ref{dH-BC}). This makes the the study of the TM modes much more complicated than that of the TE modes.

For $z\notin[0,L]$, Eqs.~(\ref{E-TM}), (\ref{H-TM}), (\ref{H-BC}), and (\ref{dH-BC}) decouple and we have plane wave solutions. For $z\in[0,L]$, we can decouple the first of these equations from the rest. To do this we use (\ref{E-TM}) to compute $|\vec E|^2$ and use (\ref{e1}) to show that $\sigma|\vec E|^2$ is a root of the cubic polynomial:
    \[p(x):=x^3+2\RE(\fn^2)\,x^2+|\fn|^4x-\sigma\, Q(z),\]
where
    \be
    Q(z):=\frac{Z_0^2}{k^2}
    \left[|\sH'(z)|^2+k^2\sin^2\theta|\sH(z)|^2\right].
    \label{Q=}
    \ee
Among the three roots of $p(x)$, only one vanishes for $\sigma=0$. Identifying $\sigma|\vec E|^2$ with this root, we have
    \be
    |\vec E(\vec r)|^2=
    \frac{1}{3\,\sigma}\left\{S(z)^{1/3}+[4\RE(\fn^2)^2-3|\fn|^4]S(z)^{-1/3}-\RE(\fn^2)\right\},
    \label{sol=}
    \ee
where
    \bea
    S(z)&:=&\frac{1}{2} \left[\Sigma(z)+\sqrt{\Delta(z)}\right],\nn\\
    \Sigma(z)&:=&27 \sigma\, Q(z) + 18 |\fn|^4 \RE(\fn^2) - 16 \RE(\fn^2)^3,\nn\\
    \Delta(z)&:=&\Sigma(z)^2+4 \Big[3 |\fn|^4 - 4 \RE(\fn^2)^2\Big]^3.\nn
    \eea
With the help of (\ref{e1}) and (\ref{sol=}), we can express $\hat\epsilon(z)$ and the matching conditions (\ref{dH-BC}) solely in terms of $\sH'(0^\pm)$ and $\sH'(L^\pm)$.

We can also determine the relevant root of $p(x)$ using perturbation theory. This yields the following expansion of the right-hand side of (\ref{sol=}) in powers of $\sigma$.
    \be
    |\vec E(\vec r)|^2=\frac{Q(z)}{|\fn|^4}-\frac{2\RE(\fn^2)Q(z)\sigma}{|\fn|^6}+\cO(\sigma^2),
    \label{pert-1}
    \ee
where $\cO(\sigma^\ell)$ stands for terms of order $\ell$ and higher in powers of $\sigma$.

In terms of the variables (\ref{scaled}) we can express the solution of the Helmholtz equation (\ref{HE-TM}) in the form
    \be
    \sH(L\bz)=\left\{\begin{array}{ccc}
    A_{-}e^{i\fK\,  {\bz}}+B_{-}e^{-i\fK\, {\bz}}&{\rm for}& {\bz}< 0,\\[3pt]
    \zeta(\bz)&{\rm for}& {\bz}\in [0,1],\\[3pt]
    A_{+}e^{i\fK\, {\bz}}+B_{+}e^{-i\fK\, {\bz}}&{\rm for}& {\bz}> 1,
    \end{array}\right.
    \label{e2-TM}
    \ee
where $\zeta:[0,1]\to\C$ satisfies
    \begin{align}
    &\zeta''(\bz)+\fK^2\tilde\fn^2\zeta(\bz)=\gamma\,F(\bz) ,
    \label{HE-TM1}
    \end{align}
$F:[0,1]\to\C$ is given by
    \bea
    F(\bz)&:=&\sec^2\theta f(\bz)\zeta(\bz)-
    \frac{\cos^2\theta f'(\bz)\zeta'(\bz)}{\fn^2\fK^2-\gamma \cos^2\theta f(\bz)},
    \label{F-def}\\
    f(\bz)&:=&|\vec E(L\bz)|^2=f_0(\bz)+\cO(\gamma),
    \label{f-def-1}\\
    f_0(\bz)&:=&\frac{Q(L\bz)}{|\fn|^4}=
    \frac{Z_0^2\left[\cos^2\theta|\zeta'(\bz)|^2+\sin^2\theta \fK^2|\zeta(\bz)|^2\right]}{\fK^2|\fn|^4},
    \label{f-def-2}
    \eea
and we have employed Eqs.~(\ref{Q=}) -- (\ref{pert-1}).

In view of (\ref{H-BC}), (\ref{dH-BC}), and (\ref{e2-TM}), $\zeta$ fulfills the boundary conditions:
	\bea
	\zeta(0)&=&A_{-}+B_{-},
	\label{BC-0=}\\
	 \zeta'(0)&=&i\fK\,\hat\epsilon(0)(A_--B_-),
	\label{BC-d0=}\\
	\zeta(1)&=&A_+e^{i\fK}+B_+e^{-i\fK},
	\label{BC-1=}\\
	 \zeta'(1)&=&i\fK\, \hat\epsilon(L) (A_+e^{i\fK}-B_+e^{-i\fK}).
	\label{BC-d1=}
	\eea
The $\hat\epsilon(0)$ and $\hat\epsilon(L)$ appearing in these equations are to be determined using  (\ref{e1}), (\ref{sol=}), and  (\ref{f-def-1}). These give
	\be
	\begin{aligned}
	\hat\epsilon(L\bz)&=
	\fn^2- \cos^2\theta\fK^{-2} f(\bz)\gamma\\
	&=\fn^2- \cos^2\theta\fK^{-2} f_0(\bz)\gamma+\cO(\gamma^2).
	\end{aligned}
	\label{ep=f}
	\ee
In particular, with the help of  (\ref{Q=}) -- (\ref{e2-TM}) and (\ref{ep=f}), we can express $\hat\epsilon(0)$ and $\hat\epsilon(L)$ in terms of $\zeta(0), \zeta'(0), \zeta(1)$, and $\zeta'(1)$. Substituting these relations together with (\ref{BC-0=}) and (\ref{BC-1=}) in (\ref{BC-d0=}) and (\ref{BC-d1=}), we obtain a pair of decoupled nonlinear equations for $\zeta'(0)$ and $\zeta'(1)$ that we can, in principle, solve to express $\zeta'(0)$  and $\zeta'(1)$ in terms of $(A_-,B_-)$ and $(A_+,B_+)$, respectively.

\subsection{Spectral Singularities and Laser Output Intensity}

According to (\ref{e2-TM}), for a left-incident TM wave with outgoing complex amplitude $N_+$,
	\begin{align}
	&B_+=0,  &&A_+=N_+,
	\label{left-}
	\end{align}
and the reflection and transmission amplitudes are given by
	\begin{align}
	&R^{\rm l}:=\frac{B_-}{A_-}, && T^{\rm l}:=\frac{N_+}{A_-}.
	\label{left-RT}
	\end{align}
Using these relations to express $A_-$ and $B_-$ in terms of $N_+$, $R^{\rm l}$, and $T^{\rm l}$, substituting the result together with (\ref{left-}) in (\ref{BC-0=}) and (\ref{BC-d0=}), and solving them for $R^{\rm l}$ and $T^{\rm l}$, we recover (\ref{R-T}) with
	\be
	G_\pm(\fK):=\zeta'(0)\pm i\fK\,\hat\epsilon(0)\zeta(0).
	\label{G-TM}
	\ee
In particular, nonlinear spectral singularities for left-incident TM waves are again given by $G_+(\fK)=0$. Following the approach of Sec.~\ref{TE} we seek a first-order perturbative solution of this equation where $\gamma$ plays the role of the perturbation parameter. This requires expressing $\zeta'(0)$ and $\zeta'(1)$ in terms of $N_+$. To do this we treat $\bz=1$ as the initial value of $\bz$ and
use first-order perturbation theory to solve the initial-value problem defined by Helmholtz equation (\ref{HE-TM1}) and the boundary conditions (\ref{BC-1=}) and (\ref{BC-d1=}), which in this context play the role of initial conditions. In view of (\ref{f-def-2}), (\ref{ep=f}), and (\ref{left-}), they take the form:
	\begin{align}
	\zeta(1)=&N_+ e^{i\fK},
	\label{BC-1=2}\\
	\zeta'(1)=&i\fK\,N_+e^{i\fK}\left(\fn^2+a|N_+|^2\gamma\right)+\cO(\gamma^2),
	\label{BC-d1=2}
	\end{align}
where
    \bea
	a&:=&\frac{-Z_0^2\cos^4\theta}{\fK^2}\left(1+\frac{\tan^2\theta}{|\fn|^4}\right).
	\label{a=3}
	\eea

Next, we apply first-order perturbation theory to determine the solution of the nonlinear Helmholtz equation (\ref{HE-TM1}) that satisfies (\ref{BC-1=2}) and (\ref{BC-d1=2}). First, we note that this solution satisfies
	\be
	\zeta(\bz)=\zeta_0(\bz)+\frac{\gamma}{\fK\,\tilde\fn}
    \int_1^\bz \sin[\fK\,\tilde\fn(\bz-\bz')]F(\bz')d\bz',
    \label{sol-01}
	\ee
where $\zeta_0$ is the solution of the linear Helmholtz equation, $\zeta''(\bz)+\fK^2\tilde\fn^2\zeta(\bz)=0$, that fulfils (\ref{BC-1=2}) and (\ref{BC-d1=2}). It is easy to show that
    \be
    \zeta_0(\bz)=\zeta_0^{(0)}(\bz)+\zeta_0^{(1)}(\bz)\gamma,
    \label{zeta-zero=}
    \ee
where
    \begin{align}
    &\zeta_0^{(0)}(\bz):= N_+ e^{i\fK}
    \left[\rC(\tilde\fn,\bz)-i \fu^{-1}\rS(\tilde\fn,\bz)\right],
    \label{zeta-00}\\
    &\zeta_0^{(1)}(\bz):=- i N_+ |N_+|^2 e^{i\fK} \tilde\fn^{-1}a\, \rS(\tilde\fn,\bz),
    \label{zeta-01}\\
    &\rC(\fz,\bz):=\cos[\fz\fK(1-\bz)],~~~~~~\fu:=\frac{\tilde\fn}{\fn^2},
    \label{rC=}\\
    &\rS(\fz,\bz):=\sin[\fz\fK(1-\bz)].
    \label{rS}
    \end{align}

Using $\zeta_0(\bz')$ for $\zeta(\bz')$ in the expression for $F(\bz')$ that appears on the right-hand side of (\ref{sol-01}) and expanding the result in powers of $\gamma$ give
    \be
    \zeta(\bz)=\zeta_0^{(0)}(\bz)+
    \left[\zeta_0^{(1)}(\bz)+\zeta_1^{(1)}(\bz)\right]\gamma+\cO(\gamma^2),
    \label{sol-pert}
    \ee
where
    \begin{align}
    &\zeta_1^{(1)}(\bz):=\frac{1}{\fK\,\tilde\fn}\int_1^z\sin[\fK\tilde\fn(\bz-\bz')]F_0(\bz')dz',
    \label{z11=}\\
    &F_0(\bz):=\sec^2\theta f_0(\bz)\zeta_0^{(0)}(\bz)-\frac{\cos^2\theta f_0'(\bz)\zeta_0^{(0)\prime}(\bz)}{\fK^2\fn^2},
    \label{F=F-def}
    \end{align}
and $f_0(\bz)$ is to be computed by substituting $\zeta_0^{(0)}$ for $\zeta$ in (\ref{f-def-2}).

Next, we use (\ref{G-TM}) and (\ref{sol-pert}) to expand $G_\pm(\fK)$ in powers of $\gamma$. In view of (\ref{ep=f}), this gives
    \be
    G_\pm(\fK)=G_{0\pm}^{(0)}(\fK,\fn)+\sum_{j=0}^1
    G_{j\pm}^{(1)}(\fK,\fn)\gamma+\cO(\gamma^2),
    \label{G-NL=}
    \ee
where
    \bea
    G_{j\pm}^{(j)}(\fK,\fn)&:=&\zeta_j^{(j)\prime}(0)
    \pm i\fK\,\fn^2\zeta_j^{(j)}(0),~~~~~j=0,1,
    \label{Gjj=}\\
    G_{0\pm}^{(1)}(\fK,\fn)&:=&\zeta_0^{(1)\prime}(0)
    \pm i\fK\,\fn^2\left[\zeta_0^{(1)}(0)-\frac{\cos^2\theta f_0(0)\zeta_0^{(0)}(0)}{\fK^2\fn^2}
    \right].
    \label{G01=}
    \eea

We can use (\ref{zeta-00}), (\ref{Gjj=}), and (\ref{G01=}) to show that
    \bea
    G_{0+}^{(0)}(\fK,\fn)&=&\frac{iN_+ e^{i(\tilde\fn+1)\fK}\fK\,\fn^2(\fu+1)^2}{2\fu}\,L(\fK,\tilde \fn,\fu),
    \label{G00=11}
    \eea
where
    \be
    L(\fK,\tilde\fn,\fu):=e^{-2i\fK\tilde n}-\left(\frac{\fu-1}{\fu+1}\right)^2.
    \label{L=}
    \ee
The linear spectral singularities correspond to the values $(\fn_0,\fK_0)$ of $(\fn,\fK)$ that fulfill $G_{0+}^{(0)}(\fK_0,\fn_0)=0$, equivalently $L(\fK_0,\tilde\fn_0,\fu_0)=0$, where $\fu_0:=\tilde\fn_0/\fn_0^2$. For the typical gain media, which satisfy (\ref{typical}),  $L(\fK_0,\tilde\fn_0,\fu_0)=0$ reduces to
    \be
    e^{-i\fK_0\tilde \fn_0}= \frac{\fu_0-1}{\fu_0+1}.
    \label{SS-L-TM}
    \ee
In analogy with the TE modes, we can use this relation to determine the threshold gain for the TM modes. This is done in Ref.~\cite{pra-2015a}. Here we quote the result:
    \be
    g_0=\frac{2\,\IM(\fn_0)}{L\cos\theta\,\IM(\tilde\fn_0)}
    \ln\left|\frac{\tilde\fn_0+\fn_0^2}{\tilde\fn_0-\fn_0^2}\right|.
    \label{th-g-TM=1}
    \ee
The wavenumber $k_0$ of the emitted radiation turns out to be given by (\ref{th-lambda=1}) provided that we identify $\varphi_0$ with the phase angle of $(\tilde\fn_0-\fn_0^2)/(\tilde\fn_0+\fn_0^2)$.

Again, we can obtain more explicit expressions for $g_0$ and $k_0$, if the gain medium fulfils (\ref{typical}). In this case, $k_0$ takes the form (\ref{th-lambda=1}) and $g_0$ reads
    \bea
    g_0&\approx&\frac{2\sqrt{\eta_0^2-\sin^2\theta}}{L\eta_0}\ln\left|
    \frac{\sqrt{\eta_0^2-\sin^2\theta}+\eta_0^2\cos\theta}{
    \sqrt{\eta_0^2-\sin^2\theta}-\eta_0^2\cos\theta}\right|.
    \label{g-M=}
    \eea
It is important to note that this relation holds only for the values of $\theta$ that are not too close to the Brewster's angle, $\theta_b:=\tan^{-1}(\eta_0)$. In the vicinity of $\theta_b$, the right-hand side of (\ref{g-M=}) takes arbitrarily large values. This violates (\ref{typical}) because, according to (\ref{g-zero}), $\kappa_0$ is proportional to $g_0$. Therefore, the approximation scheme leading to (\ref{g-M=}) fails \cite{pra-2015a}.

To characterize nonlinear spectral singularities to the first order in $\gamma$, we suppose that
    \begin{align}
    &\fK=\fK_0+\fK_1\gamma,
    &&\fn=\fn_0+\fn_1\gamma,
    \label{n1-k1=2}
    \end{align}
where $\fK_1$ and $\fn_1$ are respectively $\gamma$-independent real and complex parameters. Inserting (\ref{n1-k1=2}) in (\ref{G-NL=}), expanding the result in powers of $\gamma$, neglecting the quadratic and higher order terms, enforcing $G_+(\fK)=0$, and making use of (\ref{Gjj=}) -- (\ref{SS-L-TM}), we arrive at (\ref{NL-SS-TE-1}) with
    \bea
    \fa&:=&-\frac{\partial_{\fn_0}G_{0+}^{(0)}(\fK_0,\fn_0)}{
    \partial_{\fK_0}G_{0+}^{(0)}(\fK_0,\fn_0)},
    \label{a-TM}
    \\
    \fb&:=&-\frac{2[G_{0+}^{(1)}(\fK_0,\fn_0)+G_{1+}^{(1)}(\fK_0,\fn_0)]}{|N_+|^2
    \partial_{\fK_0}G_{0+}^{(0)}(\fK_0,\fn_0)}.
    \label{b-TM}
    \eea
It is not difficult to see that $\fa$ and $\fb$ are independent of $|N_+|^2$. In particular, if we assume that the presence of the nonlinearity does not affect the real part of the refractive index, i.e., set $\eta_1=0$, we find that its intensity has the form (\ref{eq1}) with $\widehat I$ given by (\ref{I-TE=}). The only difference is that we must now use (\ref{a-TM}) and (\ref{b-TM}) to compute the right-hand side of (\ref{I-TE=}).

Next, we try to obtain more explicit expressions for $\fa$ and $\fb$. To determine $\fa$, we insert (\ref{G00=11}) in (\ref{a-TM}) and make use of (\ref{SS-L-TM}) and the definition of $\tilde\fn_0$ and $\fu_0$ to write the result in the form:
    \be
    \fa=\frac{-\fn_0}{\fn_0^2-\sin^2\theta}\left[\fK_0-
    \frac{2i(\fn_0^2-2\sin^2\theta)}{(\fn_0^2-1)(\fn_0^2-\tan^2\theta)}\right].
    \label{fa=}
    \ee
The calculation of $\fb$ involves the use of a series of tricks that we outline in the appendix. The result is
    \be
    \fb=\frac{\fc}{\fK_0\tilde\fn_0^2}=
    \frac{\fc\,\cos^2\theta}{\fK_0(\fn_0^2-\sin^2\theta)},
    \label{fb=}
    \ee
where $\fc$ is a function of $\fK_0$, $\fn_0$, and $\theta$. We give its explicit expression in terms of $\fK_0$, $\fn_0$, and $\tilde\fn_0$ in the appendix. See Eq.~(\ref{fc=2}) below.

Having determined the parameters $\fa$ and $\fb$ for the TM modes we can obtain the laser output intensity in these modes using (\ref{eq1}) and (\ref{I-TE=}), where $\fa$ and $\fb$ are respectively given by (\ref{fa=}) and (\ref{fb=}). In order to elucidate the physical meaning of the result, we confine our attention to gain material that satisfies (\ref{typical}). This allows us to neglect terms of order $|\kappa_0|$ and $(k_0L)^{-1}$, but not of $k_0L\kappa_0$, and provides the following approximate expression that is reliable  for values of $\theta$ that are not in a close vicinity of Brewster's angle $\theta_b$.
    \be
    \widehat I\approx\frac{128 \eta_0^6 (\eta_0^2-1) \cos\theta
    \left[\eta_0^2- (\eta_0^2+1)\sin^2\theta\right]
    (\eta_0^2-\sin^2\theta)^{3/2}
    \cX(\eta_0,\theta)}{Z_0^2\sum_{\ell=0}^{5}\cY_\ell(\eta_0)\cos(2\ell\theta)},
    \label{int-TM=}
    \ee
where
    \begin{align*}
    &\cX(\eta_0,\theta):=\ln\left|\frac{\sqrt{\eta_0^2-\sin^2\theta}+\eta_0^2\cos\theta}{
    \sqrt{\eta_0^2-\sin^2\theta}-\eta_0^2\cos\theta}
    \right|,\\
    &\cY_0(\eta_0):=-14 + 55 \eta_0^2 - 82 \eta_0^4 + 101 \eta_0^6 - 132 \eta_0^8 + 40 \eta_0^{10},\\
    &\cY_1(\eta_0):= 2 (7 - 22 \eta_0^2 + 10 \eta_0^4 + 24 \eta_0^6 - 83 \eta_0^8 + 8 \eta_0^{10}),\\
    &\cY_2(\eta_0):= 4 (2 - 11 \eta_0^2 + 29 \eta_0^4 - 39 \eta_0^6 + \eta_0^8 - 6 \eta_0^{10}),\\
    &\cY_3(\eta_0):= -13 + 44 \eta_0^2 - 53 \eta_0^4 + 6 \eta_0^8,\\
    &\cY_4(\eta_0):= 6 - 11 \eta_0^2 - 2 \eta_0^4 + 7 \eta_0^6,~~~~~
    \cY_5(\eta_0):=\eta_0^4-1.
    \end{align*}
For a sample with $\eta_0=3.4$, we have compared the exact (numerical) values of $\widehat I$ with those given by (\ref{int-TM=}) and found that the difference is smaller than $10^{-3}$.

\begin{figure}
	\begin{center}
	\includegraphics[scale=.5]{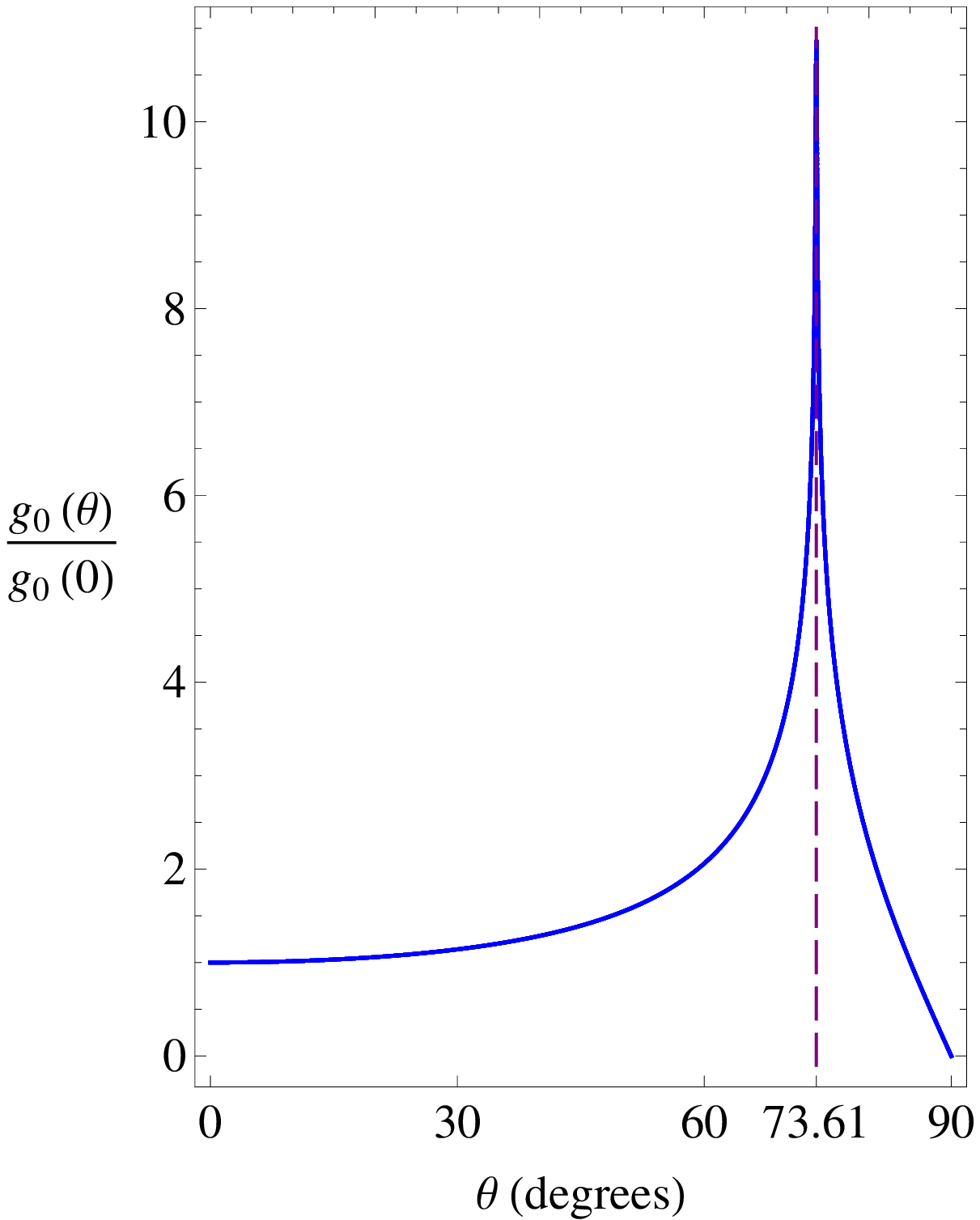}\hspace{2cm}
    \includegraphics[scale=.5]{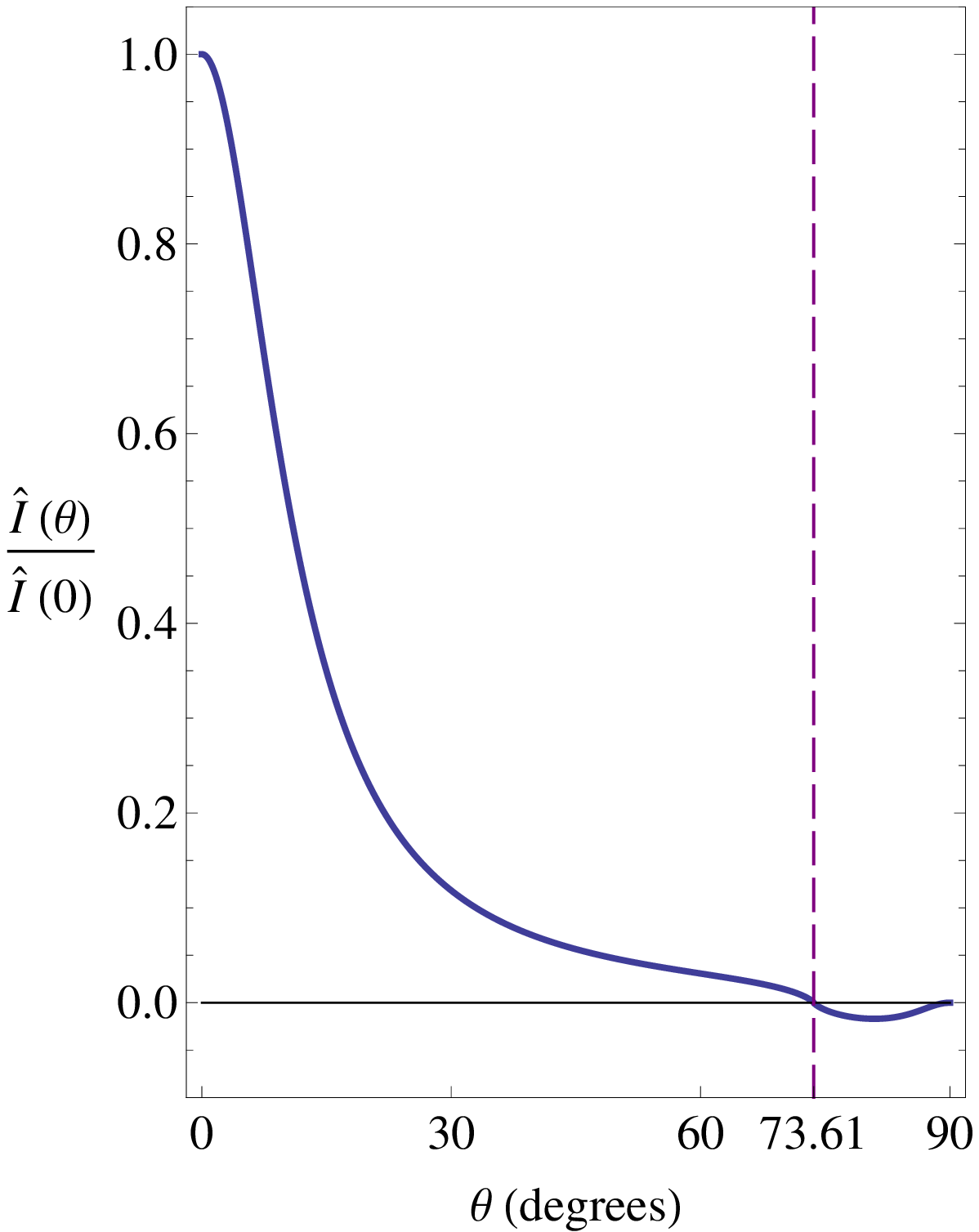}
	\caption{Plot of $g_0(\theta)/g_0(0^\circ)$ (on the left) and $\widehat I(\theta)/\widehat I(0^\circ)$ (on the right) for TM modes obtained for $L=300~\mu{\rm m}$, $\lambda_0=1500~{\rm nm}$, and $\eta_0=3.4$. Here $g_0(0^\circ)=40.409~{\rm cm}^{-1}$ and $\widehat I(0^\circ)=393.088/Z_0^2$. The dashed (purple) line marks the Brewster's angle $\theta_b$ whose value is $73.61^\circ$. For $\theta=\theta_b$, $g_0$ takes its maximum value namely $461.113~{\rm cm}^{-1}$. For $\theta\geq\theta_b$, $\widehat I(\theta)\leq 0$. This is an indication that the slab does not lase in its TM modes with $\theta\geq\theta_b$, even if the gain coefficient exceeds the threshold gain.}
	\label{fig3}
	\end{center}
	\end{figure}

Figure~\ref{fig3} shows the plots of the threshold gain and the normalized intensity factor (\ref{int-TM=}) as a function of $\theta$ for this sample.
It reveals the curious fact that for $\theta\geq\theta_b$,   $\widehat I$ takes negative values. This observation suggests that a mirrorless slab cannot emit TM-polarized laser light having $\theta\geq\theta_b$ even if one fulfills the laser threshold condition ($g>g_0$). Therefore the counterintuitive behavior of the singular TM waves with $\theta\geq\theta_b$ that is reported in \cite{pra-2015a}, do not seem to have any physical implications.

For $\theta=0$, (\ref{int-TM=}) gives $\widehat I\approx \widehat I^{\,\rm TE}/Z_0^2$, where $\widehat I^{\,\rm TE}$ is the intensity factor for the normally-incident TE mode that is given by (\ref{I-zero}). The fact that for $\eta_0\leq 3$, this quantity diverges for some $\theta_\star\geq 0$ is a sign that the assumption, $\eta_1=0$, is not valid. The situation is similar to the case of TE waves. Postulating the linear relation (\ref{eta-1-beta}) between $\eta_1$ and $\fK_1$ resolves the difficulty with the divergence and negative values of $\widehat I$ for $\theta\leq\theta_b$. It does not however affect the negativity of $\widehat I$ for $\theta>\theta_b$.

\section{Concluding Remarks}

The output intensity of a homogeneous slab laser is known to be a linear function of the gain coefficient. In this article we offer a derivation of this relation and obtain the explicit form of the slope of the intensity for arbitrary TE and TM modes of a mirrorless slab laser. Our approach relies on two basic postulates:
    \begin{itemize}
    \item[ ({i})] The emitted laser light is purely outgoing.
    \item[({ii})] The interaction of the wave with the slab is described in terms of a weak Kerr nonlinearity.
    \end{itemize}
Mathematically, these correspond to the emergence of a nonlinear spectral singularities \cite{prl-2009,prl-2013}. Imposing ({i}) in the absence of nonlinearity gives the laser threshold condition for the system. Demanding that it holds in the presence of a weak Kerr nonlinearity, i.e., enforcing (i) and (ii), allows for an analytic calculation of the laser output intensity $I$. The effective description of the emission of the laser light that is provided by (i) and (ii) encodes all the underlying microscopic physical phenomena in a single small free parameter, namely the Kerr coefficient.

The linear dependence of $I$ on the gain coefficient follows from the observation that a first-order perturbation of the wavenumber, $\gamma k_1$, induces a change in the real part of the refractive $\eta_0$ that is proportional to $\gamma k_1$ or equivalently $\gamma\fK_1$. If the proportionality constant $\beta$ is smaller than a critical value $\beta_\star$, our analysis leads to a negative output intensity for sufficiently small emission angles $\theta$. Furthermore, $\beta_\star$ has the same sign as $3-\eta_0$. For a mirrorless slab, we can fulfill the laser threshold condition provided that we use high-gain material which have a typically large refractive index. For $\eta_0\geq 1.5$, $\beta_\star$ turns out to be at most of the order of $10^{-4}$. Therefore a very mild $k_1$-dependence of the real part of the refractive index ensures a consistent application of our effective description of laser emission in both the TE and TM modes of the slab.

Our investigation reveals a drastic difference between the TE and TM modes. For the TM modes the laser output intensity vanishes at the Brewster's angle $\theta_b$ and becomes negative for $\theta>\theta_b$. This suggest that lasing in the TM modes of a mirrorless homogeneous slab is forbidden for $\theta\geq\theta_b$. This is an experimentally checkable theoretical prediction that follows as a logical consequence of the postulates (i) and (ii).

\subsection*{Acknowledgments} This work has been supported by  the Scientific and Technological Research Council of Turkey (T\"UB\.{I}TAK) in the framework of the project no: 114F357, and by the Turkish Academy of Sciences (T\"UBA).

\section*{Appendix: Calculation of $\fb$ for the TM modes}

According to (\ref{b-TM}), the calculation of $\fb$ for the TM modes requires the computation of
$G_{0+}^{(1)}(\fK_0,\fn_0)$, $G_{1+}^{(1)}(\fK_0,\fn_0)$, and $\partial_{\fK_0}G_{0+}^{(1)}(\fK_0,\fn_0)$. To achieve this, first we compute $f_0(\bz)$ for
$\zeta(\fz)=\zeta_0^{(0)}(\bz)$, $\fn=\fn_0$, and $\fK=\fK_0$. For these values of $\fn$ and $\fK$, we can use (\ref{zeta-00}) and (\ref{SS-L-TM}) to show that
    \bea
    \zeta_0^{(0)}(\bz)&=&\fu_0^{-1}N_+ e^{i\fK_0}\left[\xi_+(\bz)+\xi_-(\bz)\right],
    \label{Z00=101}\\
    \zeta_0^{(0)\prime}(\bz)&=& i\fK_0\tilde\fn_0\fu_0^{-1}
    N_+ e^{i\fK_0} \left[\xi_+(\bz)-\xi_-(\bz)\right],
    \label{dZ00=101}
    \eea
where
    \be
    \xi_\pm(\bz):=\frac{1}{2}(\fu_0\pm1)^\bz(\fu_0\mp 1)^{1-\bz}.
    \label{xi=}
    \ee
If we set $\zeta(\bz)=\zeta_0^{(0)}(\bz)$ in (\ref{f-def-2}) and use (\ref{Z00=101}) and (\ref{dZ00=101}) to simplify the resulting expression for $f_0(\bz)$, we find
    \be
    f_0(\bz)=|N_+|^2\cos^2\theta\,\widehat{f_0}(\bz),
    \label{f-def-4}
    \ee
where
    \be
    \widehat f_0(\bz):=Z_0^2\left[\left|\xi_+(\bz)-\xi_-(\bz)\right|^2+
    \frac{\tan^2\theta}{|\tilde\fn_0|^2}\left|\xi_+(\bz)+\xi_-(\bz)\right|^2\right]
    \label{wide-f0=}
    \ee
The following relations are straightforward consequences of (\ref{Z00=101}) -- (\ref{wide-f0=}).
    \bea
    \zeta_0^{(0)}(0)&=&\zeta_0^{(0)}(1)=N_+ e^{i\fK_0},
    \label{app-b1}\\
    \zeta_0^{(0)\prime}(0)&=&-\zeta_0^{(0)\prime}(1)=-i\fn_0^2\fK_0 N_+ e^{i\fK_0}.
    \label{app-b2}\\
    \widehat{f_0}(0)&=&\widehat{f_0}(1)=Z_0^2\left(1+\frac{\tan^2\theta}{|\fn_0|^4}\right)=
    -\sec^4\theta\fK_0^2a,
    \label{f-zero-one}
    \eea
where $a$ is defined by (\ref{a=3}). Now, we substitute (\ref{app-b1}) -- (\ref{f-zero-one}) in (\ref{G01=}) to obtain
    \be
    G_{0+}^{(1)}(\fK_0,\fn_0)=2i a \fK_0|N_+|^2N_+ e^{i\fK_0}.
    \label{G01=comp}
    \ee

The computation of $G_{1+}^{(1)}(\fK_0,\fn_0)$ is much more complicated. First, we use
(\ref{zeta-00}), (\ref{z11=}), and (\ref{Gjj=}) to show that
    \be
    G_{1+}^{(1)}(\fK_0,\fn_0)=-\frac{e^{-i\fK_0}}{N_+}\int_0^1\zeta_0^{(0)}(1-\bz)F_0(\bz)d\bz.
    \label{G11=id}
    \ee
In view of (\ref{F=F-def}), the integrand in (\ref{G11=id}) involves terms of the form $\zeta_0^{(0)}(1-\bz)\zeta_0^{(0)}(\bz)$ and $\zeta_0^{(0)}(1-\bz)\zeta_0^{(0)\prime}(\bz)$. By virtue of (\ref{SS-L-TM}), these take the following simple form.
    \bea
    \zeta_0^{(0)}(1-\bz)\zeta_0^{(0)}(\bz)&=&
    \left(\frac{\fu_0^2-1}{2\fu_0^2}\right)N_+^2e^{2i\fK_0}
    \big[{\rm C}(\tilde\fn_0,2\bz)+1\big],
    \label{id1=}\\
    \zeta_0^{(0)}(1-\bz)\zeta_0^{(0)\prime}(\bz)&=&
    \left(\frac{\fu_0^2-1}{2\fu_0^2}\right)
    N_+^2e^{2i\fK_0}\fK_0\tilde\fn_0 {\rm S}(\tilde\fn_0,2\bz),
    \label{id2=}
    \eea
where in the definition of ${\rm C}$ and ${\rm S}$, i.e., (\ref{rC=}) and (\ref{rS}), we set $\fK=\fK_0$.

Next, we substitute (\ref{F=F-def}) in (\ref{G11=id}), make use of (\ref{f-def-4}), (\ref{id1=}), and (\ref{id2=}), and perform an integration by parts to obtain
    \bea
    G_{1+}^{(1)}(\fK_0,\fn_0)&=&-G_{0+}^{(1)}(\fK_0,\fn_0)-\frac{1}{2}\,N_+|N_+|^2e^{i\fK_0}
    (1-\fu_0^{-2})\fc,
    \label{G11=AP-1}
    \eea
where
    \bea
    \fc:=\int_0^1
    \left[(1-2\fn_0^2\fu_0^2\cos^4\theta){\rm C}(\tilde\fn_0,2\bz)+1 \right]\widehat{f_0}(\bz)d\bz.
    \label{fc=}
    \eea
In the derivation of (\ref{G11=AP-1}) we have also made use of (\ref{f-zero-one}), (\ref{G11=id}), and the identities $\sin(\fK_0\tilde\fn_0)= -2i\fu_0/(\fu_0^2-1)$ and $\cos(\fK_0\tilde\fn_0)=(\fu_0^2+1)/(\fu_0^2-1)$, which follow from (\ref{SS-L-TM}).

Equation~(\ref{G11=AP-1}) reduces the calculation of $G_{1+}^{(1)}(\fK_0,\fn_0)$ to that of $\fc$. To do this, first we note that for $\fK=\fK_0$,
    \be
    \rC(\tilde\fn_0,2\bz)=\frac{\xi_-(2\bz)}{\fu_0-1}
    +\frac{\xi_+(2\bz)}{\fu_0+1}.
    \label{cos=2}
    \ee
Inserting (\ref{wide-f0=}) and (\ref{cos=2}) in (\ref{fc=}), evaluating the resulting integral,
and making extensive use of (\ref{SS-L-TM}), we find
    \be
    \fc=Z_0^2\left[\left(1-2\cos^4\theta\,\fn_0^{-2}\tilde\fn_0^2\right)\fc_1+\fc_0\right],
    \label{fc=2}
    \ee
where
    \begin{align}
    &\fc_j:=\fc_{j1}+\frac{\tan^2\theta \:\fc_{j2}}{|\tilde\fn_0|^2},~~~~j=0,1,
    \nn\\[3pt]
    &\fc_{01}:=
    -\frac{\RE(\tilde\fn_0^2/\fn_0^2)}{\fK_0\RE(\tilde\fn_0)\IM(\tilde\fn_0)},
    \nn\\[3pt]
    &\fc_{02}:=
    -\frac{|\tilde\fn_0|^2\RE(\fn_0^2)}{\fK_0|\fn_0|^4\RE(\tilde\fn_0)\IM(\tilde\fn_0)},
    \nn\\[3pt]
    &\fc_{11}:=-\frac{\fc_{01}}{2}-\frac{i
    \left[3\fn_0^{2*}\tilde\fn_0^2(\tilde\fn_0^2+3\fn_0^4)+
    \tilde\fn_0^{2*}\fn_0^2(3\tilde\fn_0^2+\fn_0^4)\right]}{\fK_0|\fn_0|^4
    (\tilde\fn_0^2-\fn_0^4)(9\tilde\fn_0^2-\tilde\fn_0^{2*})},
    \nn\\[3pt]
    &\fc_{12}:=\frac{\fc_{02}}{2}-\frac{i|\tilde\fn_0|^2
    \left[\fn_0^{2*}(\tilde\fn_0^2+3\fn_0^4)+
    3\fn_0^2(3\tilde\fn_0^2+\fn_0^4)\right]}{\fK_0|\fn_0|^4
    (\tilde\fn_0^2-\fn_0^4)(9\tilde\fn_0^2-\tilde\fn_0^{2*})}
    .\nn
    \end{align}

The final step of the computation of $\fb$ is the determination of $\partial_{\fK_0}G_{+0}^{(0}(\fK_0,\fn_0)$. With the help of (\ref{G00=11}) -- (\ref{SS-L-TM}), we can easily show that $\partial_{\fK_0}G_{+0}^{(0)}(\fK_0,\fn_0)=N_+ e^{i\fK_0}\fK_0\fn_0^4(\fu_0^2-1)$. Substituting this relation together with (\ref{G11=AP-1}) in (\ref{b-TM}) yields (\ref{fb=}).

\ed